\documentclass[sigconf,screen]{acmart}

\AtBeginDocument{%
  }

\usepackage{algorithmic}
\usepackage{graphicx}
\usepackage{subcaption} 
\usepackage{textcomp}
\usepackage{xcolor}
\usepackage{booktabs}
\usepackage{listings}
\usepackage{tikz}
\usetikzlibrary{arrows.meta}
\usetikzlibrary{arrows.meta,positioning,calc,backgrounds}

\usetikzlibrary{shapes, arrows, shadows, positioning}

\newboolean{showcomments}
\setboolean{showcomments}{true} 
\ifthenelse{\boolean{showcomments}}
{\newcommand{\nb}[2]{\fbox{\bfseries\sffamily\scriptsize#1}{\sf\small$\blacktriangleright$\textit{#2}$\blacktriangleleft$}}}
{\newcommand{\nb}[2]{}}

\lstdefinelanguage{YAML}{
  keywords={true,false,null,True,False,NULL},
  keywordstyle=\color{blue}\bfseries,
  basicstyle=\ttfamily\small,
  comment=[l]{\#},
  commentstyle=\color{gray},
  string=[s]{"}{"},
  morestring=[s]{'}{'}, 
  alsoletter={-},       
  sensitive=true,
}

\lstdefinestyle{yamlstyle}{
  language=YAML,
  basicstyle=\ttfamily\small,
  columns=fullflexible,
  showstringspaces=false,
  keepspaces=true,
  breaklines=true,
  frame=single,
  framerule=0.4pt,
  rulecolor=\color{black!30},
  numbers=left,
  numbersep=6pt,
  numberstyle=\tiny\color{black!50},
  tabsize=2
}

\begin{document}


\title{SPARK: Secure Predictive Autoscaling for Robust Kubernetes}

\author{Zhijun Jiang}
\affiliation{
  \institution{New York Institute of Technology}
  \city{Vancouver}
  \state{BC}
  \country{Canada}}
\email{cjiang11@nyit.edu}

\author{Amin Milani Fard}
\affiliation{
  \institution{New York Institute of Technology}
  \city{Vancouver}
  \state{BC}
  \country{Canada}}
\email{amilanif@nyit.edu}

\begin{abstract}
Achieving high availability and robust security in Kubernetes requires more than reactive scaling and standard perimeter firewalls. Traditional autoscalers, such as HPA, often fail to react quickly to traffic spikes and cannot distinguish between legitimate flash crowds and DDoS attacks. We present an open-source toolchain to provide a traffic-aware autoscaling approach that utilizes an eBPF-based networking layer to enforce security policies at the kernel level while orchestrating scaling decisions based on predictive models. Our results demonstrate that the predictive approach reduces timeout errors by 32\% during sudden traffic surges compared to standard reactive scaling, while ensuring immediate network convergence and layer 7 security isolation for newly scaled pods.
\end{abstract}




\maketitle

\section{Introduction}

Kubernetes has become the de facto standard for container orchestration, but its native Horizontal Pod Autoscaler (HPA) 
has limitations in handling event-driven workloads. Kubernetes Event-driven Autoscaling (KEDA)\footnote{\url{https://keda.sh/}} 
is an open-source solution to scaling based on external metrics. However, threshold-based scaling determined solely on reactive metrics can be inefficient. PredictKube\footnote{\url{https://dysnix.com/predictkube}} 
introduces predictive scaling using machine learning 
as a proprietary software as a service. However, it lacks granular traffic differentiation capability. Existing work, such as \cite{toka2020adaptive,toka2021machine,dang2021deep,qiu2023aware,yuan2024time},  focus primarily on performance and resource efficiency and overlook mechanisms to distinguish legitimate from malicious traffic, leaving autoscalers vulnerable to Denial of Service (DoS) (aiming at downtime) and Denial‑of‑Wallet (DoW) attack \cite{kelly2021denial} (maximize billing costs).

We present \textit{SPARK}\footnote{\url{https://github.com/nyit-vancouver/SPARK}} (Secure Predictive Autoscaling for Robust Kubernetes) as an integrated tool that learns demand patterns to predict scaling needs and uses eBPF datapath programs to capture flow/latency/packet anomalies, enforce  policies, and provide feedback signals to the autoscaler. Our work contributes to the open-source tooling for Kubernetes security by integrating KEDA with predictive autoscaling and Cilium \footnote{\url{https://cilium.io/}}, an eBPF-based Container Network Interface (CNI) for low-level network security and observation, and implementing a traffic-aware scaling policy that scales for legitimate traffic. 
Our evaluations show that predictive scaling, when protected by a legitimacy-aware eBPF shield, provides a balance of performance and security during sudden traffic surges.


\section{System Design}



Figure \ref{fig:arch_diagram} presents SPARK defense and scaling architecture:

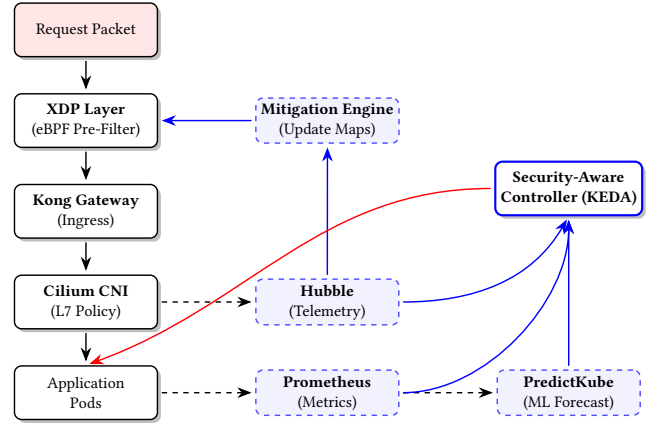
\begin{figure}[t]
\centering
\resizebox{\linewidth}{!}{
\begin{tikzpicture}[
  node distance=0.7cm and 0.8cm, 
  box/.style={rectangle, draw=black, thick, rounded corners, align=center, minimum width=2.8cm, minimum height=1.05cm, fill=white, drop shadow},
  control/.style={rectangle, draw=blue!70, thick, dashed, rounded corners, align=center, minimum width=2.8cm, fill=blue!5},
  arrow/.style={-{Stealth[scale=1.2]}, thick, shorten >=2pt, shorten <=2pt},
  bluearrow/.style={arrow, blue},
  redarrow/.style={arrow, red},
  telem/.style={arrow, dashed},
]

\node[box, fill=red!10] (edge) {Request Packet};
\node[box, below=of edge] (xdp) {\textbf{XDP Layer} \\ (eBPF Pre-Filter)};
\node[box, below=of xdp] (kong) {\textbf{Kong Gateway} \\ (Ingress)};
\node[box, below=of kong] (cilium) {\textbf{Cilium CNI} \\ (L7 Policy)};
\node[box, below=of cilium] (pods) {Application \\ Pods};

\node[control, right=of xdp, xshift=1.1cm] (mit) {\textbf{Mitigation Engine} \\ (Update Maps)};
\node[control, right=of cilium, xshift=1.1cm] (hubble) {\textbf{Hubble} \\ (Telemetry)};
\node[control, right=of pods, xshift=1.1cm] (prom) {\textbf{Prometheus} \\ (Metrics)};
\node[control, right=of prom, xshift=1.1cm] (ml) {\textbf{PredictKube} \\ (ML Forecast)};

\node[box, draw=blue, very thick, above=3.0cm of ml] (keda) {\textbf{Security-Aware} \\ \textbf{Controller (KEDA)}};

\draw[arrow] (edge) -- (xdp);
\draw[arrow] (xdp) -- (kong);
\draw[arrow] (kong) -- (cilium);
\draw[arrow] (cilium) -- (pods);

\draw[telem] (cilium.east) -| ([xshift=3mm]cilium.east) |- (hubble.west);
\draw[telem] (pods.east)   -| ([xshift=3mm]pods.east)   |- (prom.west);
\draw[telem] (prom.east)   -- (ml.west);

\draw[bluearrow] (hubble.north) to[out=90, in=-90] (mit.south);
\draw[bluearrow] (mit.west) -- (xdp.east); 

\draw[bluearrow] (prom.east) to[out=0, in=-90, looseness=0.9] (keda.south);

\draw[bluearrow] (ml.north) -- (keda.south);

\draw[bluearrow] (hubble.east) to[out=0, in=-120] (keda.south); 

\draw[redarrow] (keda.west) to[out=180, in=20] (pods.north);

\end{tikzpicture}
}
\caption{SPARK architecture: data and control planes are shown in black and blue respectively. Arrows represent traffic (black), control signals (blue), and scaling actions (red).}
\label{fig:arch_diagram}
\end{figure}

\begin{itemize}
    \item \textbf{Tier 1: XDP Pre-Filter.} Located at the network edge, the XDP program maintains a blocklist and rate-limit map in the kernel. It drops volumetric attacks (SYN floods) before OS stack processing and prevents malicious packets from generating metrics that would falsely trigger KEDA scaling. We use Kong Gateway to manage ingress traffic and improve stability under high load compared to port forwarding.
    
    \item \textbf{Tier 2: Cilium L7 Policies.} Located at the CNI layer, Cilium enforces identity-based policies by limiting HTTP rate per source identity (e.g., preventing a single compromised frontend from flooding the backend). We use Hubble \footnote{\url{https://github.com/cilium/hubble}}, 
    a networking and security observability platform built on Cilium and eBPF, to capture layer 7 flow data (HTTP status codes) to calculate the traffic legitimacy score.
    \item \textbf{Tier 3: Security-Aware Controller.} This custom component acts as a meta-scaler. It ingests Reactive Metrics (Prometheus), Proactive Forecasts (PredictKube), and Legitimacy Scores (Hubble) to reconcile conflicting signals. We deploy KEDA to enable event-driven autoscaling using reactive scaler based on HTTP request metrics with Prometheus and predictive scaler with PredictKube.
\end{itemize}






\textbf{Integration of Cilium eBPF Datapath with KEDA.} When KEDA scales and new pods join the cluster, Cilium wires them into the network data plane by updating eBPF maps in the kernel ensuring that new pods are load-balanced instantly. This provides immediate connectivity without the latency related to iptables convergence. Cilium policies can restrict access for scaled pods.




\textbf{Metrics and Telemetry Collection.} Data collection was centralized through a Prometheus and Grafana stack. We developed a custom dashboard with Grafana to visualize HTTP request rates using Prometheus metrics (\texttt{http\_requests\_total}), pod scaling events (tracked using KEDA's \texttt{ScaledObject} status), connection drops (measured using custom metrics and Grafana alerts), traffic legitimacy (ratio of status 200 to 400+ responses), and resource utilization (memory/CPU usage vs. limits). Hubble captures L7 flow data, specifically tracking dropped packets and policy verdicts.



\textbf{Distinguishing Traffic Legitimacy.} To prevent DoW attacks, we implement a traffic legitimacy filter. We define the Legitimacy Score as the sum of HTTP responses with status codes in 2xx class (successful) over total responses e.g. 2xx and 4xx. This is calculated via Hubble metrics as \textit{Legitimacy Score} = $\frac{\sum \text{HTTP}_{2xx}}{\sum \text{HTTP}_{total}}$. 
We consider traffic legitimate if the score is greater than or equal to 0.85. As a policy, the system should scale unless illegitimate traffic is detected. 
Our controller logic fuses this score with the forecast. When the Legitimacy Score is below 0.85, indicating high error rates in typical attacks, the scaling is capped regardless of the demand forecast.

    



\textbf{Predictive Scaler.} We integrate PredictKube by obtaining an API key, configuring it in KEDA, connection with PredictKube's gRPC client, and implementing prediction window tuning (5-10 min). Since this is a closed commercial product, it is difficult to interpret the reason for scaling. We are currently implementing and evaluating our open-source ML-based scaler using an LSTM neural network trained on a history of traffic fluctuation. 


\section{Evaluation}

We evaluate our approach on an Amazon EKS cluster (Kubernetes v1.27) using 4 m5.xlarge nodes, with 2 m3.medium worker nodes to support rapid scaling. We synthesize the traffic load using Vegeta, an HTTP load testing tool. To mock realistic workloads, we developed a microservice in Go v1.22 that includes a configurable startup delay for simulation of initialization to stress-test the "Scale Lag" metric. We evaluate autoscalers on two traffic profiles: (1) Flash Crowd (Legitimate) and (2) Mixed Attack Vector (Legitimate + DDoS).

\textbf{Scenario 1: Flash Crowd (Legitimate Surge).} We simulate a rapid ramp-up from 0 to 500 RPS over 30 seconds, sustained for 5 minutes, to test the scaler's ability to provision resources before requests time out. For \textbf{reactive scaling} we deployed the KEDA Prometheus Scaler with a threshold of 50 RPS per pod. The Flash Crowd traffic profile was executed, and we measured the "Time to Stabilize" and "Total Error Count" to establish a reactive baseline. For \textbf{predictive scaling} we deployed the KEDA PredictKube Scaler and trained the model using 60 minutes of historical traffic data generated by a recurring Vegeta pattern. The same Flash Crowd profile was executed to measure improvements. As shown in Table \ref{tab:results}, the predictive model significantly outperformed the reactive model. The results indicate that PredictKube reduced the Scale Lag (delay) by 54.8\%, and reduced timeouts by 32.6\%.

\begin{table}[t]
\footnotesize
\caption{Performance under flash crowd (500 RPS in 5 min).}
\label{tab:results}
\centering
\begin{tabular}{@{}lccc@{}}
\toprule
\textbf{Scaler} & \textbf{Avg. Pods} & \textbf{Timeout Rate} & \textbf{Scale Lag} \\
\midrule
Reactive & 7.2 & 18.7\% & 42s \\
Predictive & 8.1 & 12.6\% & 19s \\
\textbf{Improvement} & +12.5\% & \textbf{-32.6\%} & \textbf{-54.8\%} \\
\bottomrule
\end{tabular}
\end{table}

\textbf{Scenario 2: Mixed Traffic (Legitimate + DDoS).} We simulate a mixed traffic comprising 80\% legitimate requests (HTTP 200) and 20\% malformed malicious requests (HTTP 4xx/5xx), to test the system's ability to distinguish traffic validity. We evaluate our \textbf{security-aware scaling} by enabling the Cilium Network Policies and legitimacy gate logic. 
The reactive scaler treated all traffic as legitimate and aggressively scaled out to 15 pods, thus wasting resources. The security-aware controller detected a Legitimacy Score below 0.85 and capped scaling at 8 pods, while the XDP layer dropped 92\% of malicious traffic at the ingress. 

\section{Conclusion and Future Work}

Integrating Cilium's eBPF datapath with KEDA's event-driven autoscaling creates a robust defense against service unavailability and resource exhaustion. Our results confirm that proactive scaling, when protected by a legitimacy-aware eBPF shield, provides a balance of performance and security, reducing scale lag and timeout, as well as mitigating DDoS-induced autoscaling. 
As future work, we plan to (1) fully implement and evaluate our predictive model, (2) compare against PredictKube, (3) enhance our controller by providing explainability into scaling decisions for operators, and (4) adaptive and dynamic adjustment of the Legitimacy Score threshold based on learned baseline traffic patterns.

\begin{acks}
This work was supported by a research grant from the New York Institute of Technology - Vancouver.
\end{acks}

\bibliographystyle{ACM-Reference-Format}
\bibliography{references}

\end{document}